\documentclass{aip-cp}

\usepackage[numbers]{natbib}
\usepackage{rotating}
\usepackage{graphicx}
\usepackage{lineno}

\newcommand{\ttbar}{\ensuremath{t\overline{t}}}
\newcommand{\ttbarjet}{\ensuremath{\sigma_{t\overline{t} + 1 jet}}}
\newcommand{\mt}{\ensuremath{m_t}}
\newcommand{\mtp}{\ensuremath{m_t^{pole}}}
\newcommand{\mlb}{\ensuremath{m_{lb}}}
\newcommand{\mmlb}{\ensuremath{m_{lb}^{min}}}

\begin{document}
%\linenumbers

% Title portion
\title{Measurements of the top-quark mass in fixed schemes and with alternative methods using the ATLAS and CMS detectors at the LHC}

%\author[aff1]{Teresa Barillari~\footnote{Speaker}~\footnote{On behalf of the ATLAS and CMS collaborations} \corref{cor1}}
\author[aff1]{Teresa Barillari\corref{cor1}}
\affil[aff1]{Max-Planck-Inst. f\"{u}r Physik Werner-Heisenberg-Institut -- F\"{o}hringer Ring 6  D-80805 M\"{u}nchen, Germany}
\corresp[cor1]{barilla@mppmu.mpg.de~\footnote{Speaker}~\footnote{On behalf of the ATLAS and CMS collaborations}}

\maketitle

\begin{abstract}
Selected topics of the top-quark mass measurements 
in well-defined schemes are presented. The measurements have been 
performed using data recorded with the ATLAS and CMS 
detectors at the LHC at proton-proton centre-of-mass energies 
of 7 and 8 TeV. Precision theoretical QCD calculations 
for both inclusive top-antitop quark pair production and 
top-antitop quark pair production with an additional jet
to extract the top quark mass in the pole-mass scheme 
have been used.

\end{abstract}

% Head 1
\section{INTRODUCTION}
The top quark is by far the heaviest known fermion and the heaviest 
known fundamental particle. It plays an important role 
in the Standard Model (SM). Precise measurements of the top-quark mass
($\mt$) provide a key input to consistency tests of the SM. 
The mass of the Higgs boson and the top quark are also important 
parameters in the determination of the vacuum 
stability~\cite{Degrassi:2012ry,Baak:2014ora}.

Nowadays, the most precise determinations of $\mt$ have been achieved 
experimentally from kinematical reconstruction of the measured top-quark 
decay products, e.g. measuring the semi-leptonic decay channel of 
top-antitop quark pairs ($\ttbar$), where one top quark decays into a 
b quark, a charged lepton and its neutrino and the other top quark decays 
into a b quark and two u/d/c/s quarks, yielding a value of 
$\mt =$ 172.35 $\pm$ 0.51 GeV~\cite{Khachatryan:2015hba}.
%the di-leptonic decay channel of top-antitop 
%quark pairs ($\ttbar$), where both top quarks decay into a b quark, and 
%two opposite signed charged leptons ($t\to W b$, and $W \to l\nu$).
These $\mt$ determinations, however, have not been linked so far 
in an unambiguous manner to a Lagrangian top-quark mass in
a specific renormalization scheme 
%employed in the Standard Model.  
%lack a clear interpretation in terms 
%of a well-defined $\mt$ theoretical scheme 
as employed in perturbative calculations in quantum chromodynamics (QCD), 
electroweak fits, or any theoretical prediction in 
general~\cite{Moch:1702814,Buckley:2011ms,Hoang:2014oea}. 
The values of $\mt$ extracted using these schemes are usually
identified with the top-quark pole mass, $\mtp$. Present studies 
estimate differences between the two top-quark mass definitions, $\mt$ 
and a theoretically well defined short-distance mass definition at a 
low scale  (e.g. $\mtp$), of about 1 GeV.
%~\cite{Moch:2014tta, ATLAS:2014wva,Buckley:2011ms,Hoang:2014oea}.

In addition to direct $\mt$ measurements as mentioned above, the mass 
dependence of the QCD prediction for the cross section ($\sigma_{\ttbar}$) can be 
used to determine $\mt$ by comparing the measured to the predicted 
$\sigma_{\ttbar}$~\cite{Alekhin:2012py,Ahrens:2011px,Langenfeld:2009wd,Abazov:2011pta,Beneke:2011mq,Beneke:2012wb}. 
Although the sensitivity of $\sigma_{\ttbar}$ to $\mt$ might not be strong 
enough to make this approach competitive in precision, it yields
results affected by different sources of systematic uncertainties compared 
to the direct $\mt$ measurements and allows for extractions of $\mt$ in 
theoretically well-defined mass schemes.
The values extracted using these methods are usually identified with the 
top-quark pole mass.

This distinction of the theoretical description of the measured parameter, 
e.g. either the parameter in the underlying Monte Carlo (MC) generator, 
$m_t^{MC}$ (or simply $\mt$), the mass term in the top-quark propagator, 
$\mtp$, or the mass in a well defined low-scale short distance 
scheme~\cite{Moch:1702814,Hoang:2008xm}, is recently gaining in importance.

In the following, selected $\mtp$ measurements performed by the 
ATLAS~\cite{Aad:2008zzm} and CMS~\cite{Chatrchyan:2008aa} experiments at 
LHC~\cite{Evans:2008zzb} using data at proton-proton (pp) centre-of-mass 
energies of 7 and 8 TeV, are presented.
\section{TOP-QUARK POLE MASS MEASUREMENTS}
In contrast to the standard  kinematical reconstruction of the measured 
top-quark decay product methods mentioned above, cross-section-like 
observables can be used to compare QCD predictions depending on 
$\mtp$, with unfolded data.
The unfolding removes detector effects, and, in addition 
these measurements benefit from the larger independence from the mass 
definition in the used MC generators. 
For the total cross-section measurements, however, a 5\% uncertainty 
translates into a 1\% uncertainty in the top-quark mass~\cite{Alekhin:2013nda} 
and the difference from going from next-to-leading order (NLO) to 
next-to-next-to-leading order (NNLO) predictions is even larger 
($\sim$ 10\%).
%~\cite{Alekhin:2013nda}.
Experimentally the challenges lie in the unfolding of the data and in the 
absolute normalization. 
Furthermore measurements of $\mtp$  involving new shape-like observables as 
proposed in~\cite{Alioli:2013mxa} can help reduce both theoretical and 
experimental uncertainties. 
\subsection{Measurements of Top-Quark Pole Mass in $\ttbar$ Di-Lepton Events}
The measurements of the $\ttbar$ production cross-section, 
$\sigma_{\ttbar}$, together with the NNLO prediction 
in QCD including the resummation of next-to-next-to-leading-logarithmic 
(NNLL) soft gluon terms~\cite{Czakon:2011xx}, 
 are used to determine the top-quark pole mass.
Most of such measurements are performed in the electron-muon ($e$ - $\mu$) 
channel, where each W boson from the top quark decays into a lepton 
and a neutrino.
Events are required to contain an oppositely charged  $e$ - $\mu$ pair. 
The restriction to the di-lepton channel allows obtaining a particular 
clean $\ttbar$ event sample.
The value of $\mtp$ is determined from the $\sigma_{\ttbar}$ measurements 
in pp collisions at centre-of-mass energies of $\sqrt{s} = 7$ TeV and 
$\sqrt{s} = 8$ TeV with the CMS and ATLAS detector at LHC. 
Both experiments assume a top-quark mass of $m_t^{MC} = 172.5$ GeV 
in simulations to extract the reconstruction efficiency.
\subsubsection{CMS Top-Quark Pole Mass Measurements}
Compared to previous $\mtp$ measurements at 
7 TeV~\cite{Chatrchyan:2012bra} and at 8 TeV~\cite{Chatrchyan:2013faa} 
the latest CMS results~\cite{CMS-PAS-TOP-13-004} include the full CMS data 
samples with integrated luminosities of 5.0 $fb^{-1}$ (7 TeV) and 
19.7 $fb^{-1}$ (8 TeV). 
The value of $\mtp$ at NNLO+NNLL is extracted by confronting the measured 
cross section $\sigma_{\ttbar}$ at 7 and 8 TeV with predictions 
employing different parton density function (PDF) sets: 
NNPDF3.0~\cite{Ball:2014uwa}, 
CT14~\cite{Dulat:2015mca}, and MMHT2014~\cite{Harland-Lang:2014zoa}.
The obtained $\mtp$ values are listed in Table~\ref{tab:mpole}.
The contributions from uncertainties on the CT14 PDF set are scaled
to 68\% confidence level.

\begin{figure}[ht]
%  \centerline{\includegraphics[width=250pt]{CMS-PAS-TOP-13-004-Figure-010-a.pdf}
%              \includegraphics[width=250pt]{CMS-PAS-TOP-13-004-Figure-010-b.pdf}}
  \centerline{\includegraphics[width=220pt]{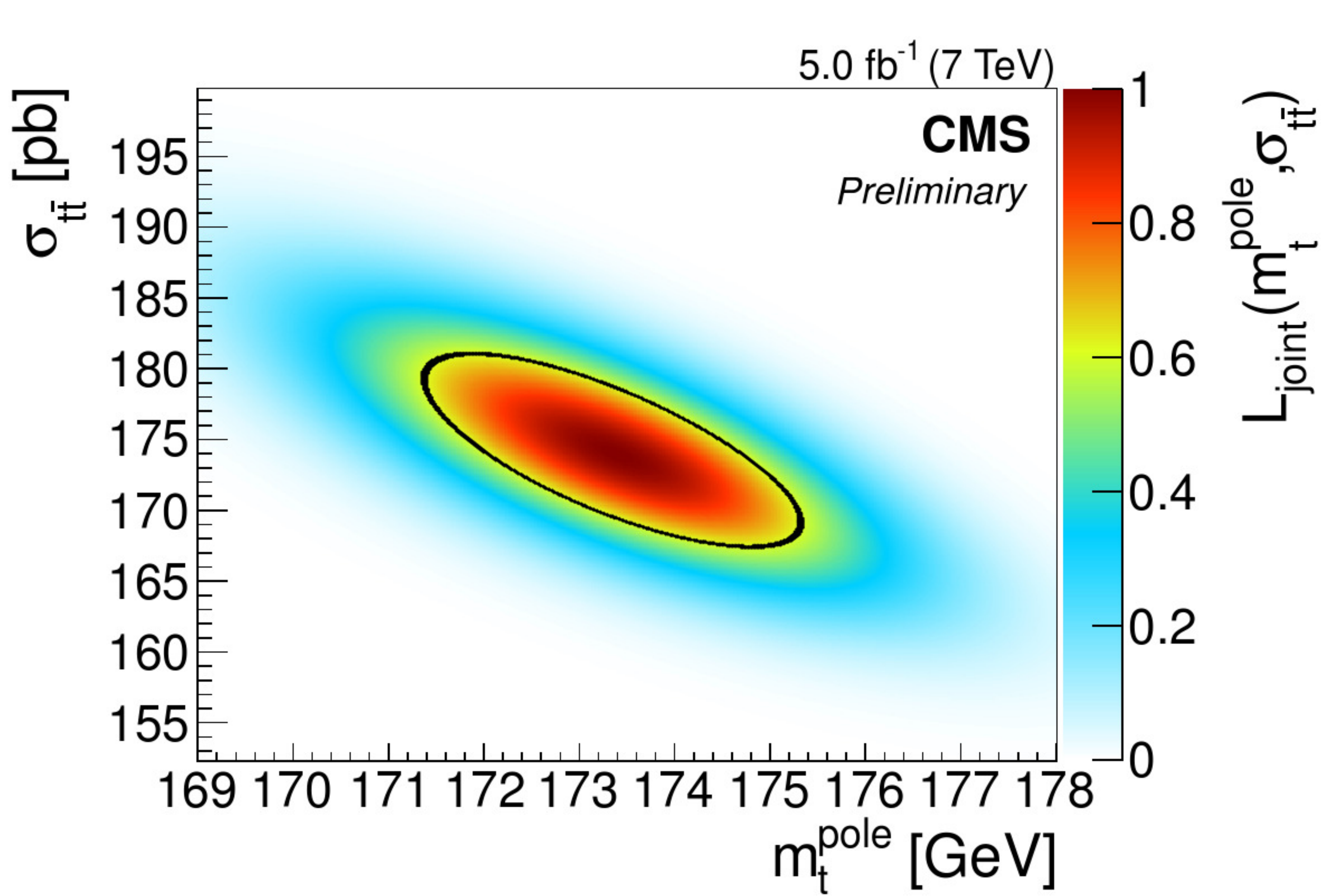}
              \includegraphics[width=220pt]{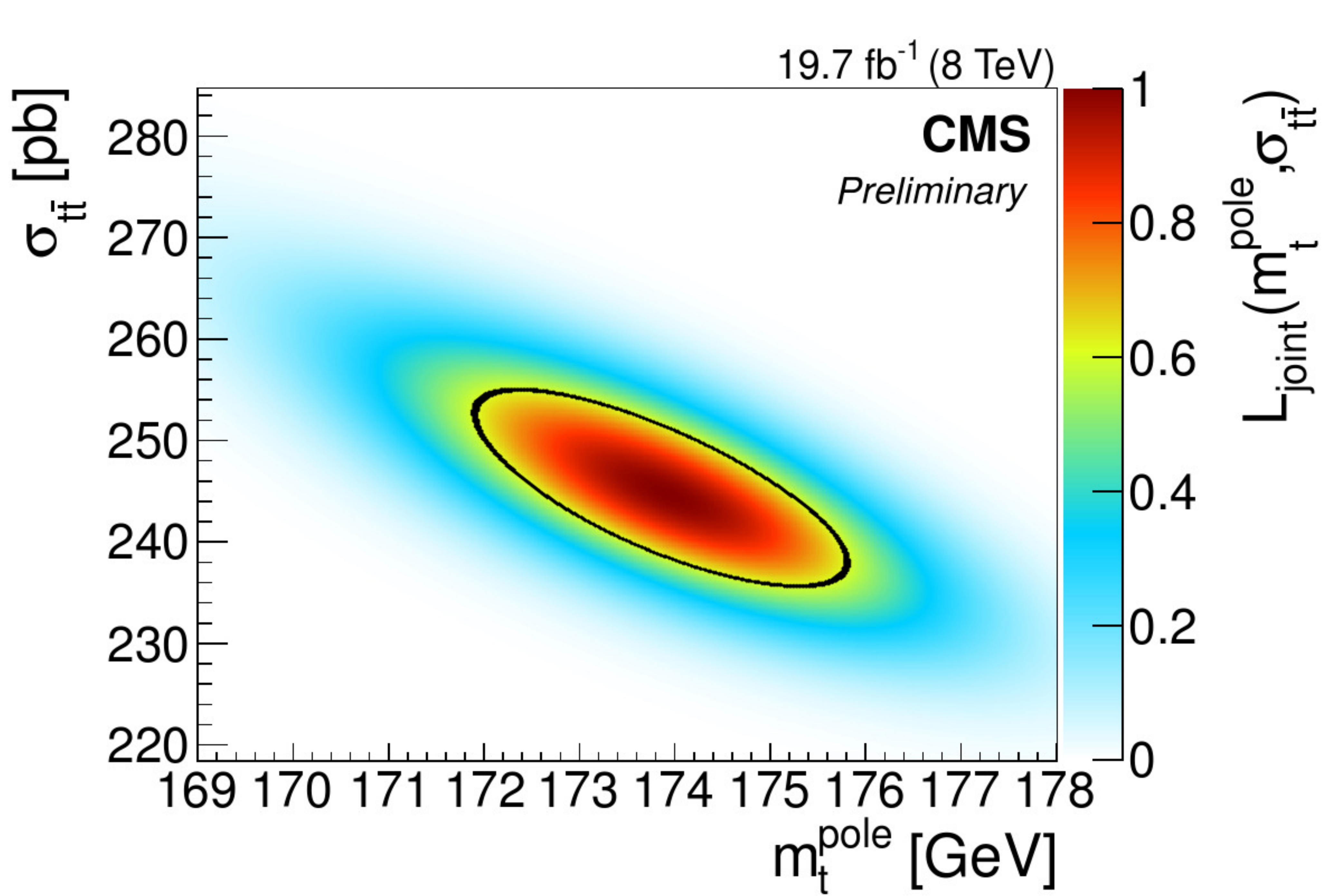}}
  \caption{Combined likelihood of the measured and predicted dependence of 
  the $\ttbar$ production cross section on the top-quark mass for 7 (left plot)
   and 8 TeV (right plot) in CMS~\cite{CMS-PAS-TOP-13-004}. The total 
   one standard deviation uncertainty is indicated by a black contour. }
%This value corresponds to data corrected for detector and hadronization effects after the background subtraction.}
\label{fig:1}
\end{figure}

A weighted average is calculated, taking into account all systematic 
uncertainty correlations between the measured cross sections at 7 
and 8 TeV and assuming 100\% correlated uncertainties
for the theory predictions at the two energies. 
The combined $\mtp$ results are listed in Table~\ref{tab:mpole}
 and are in good agreement with each other and the world average 
value~\cite{ATLAS:2014wva}.
\begin{table}[ht]
\caption{Top-quark pole mass measured by CMS at NNLO+NNLL extracted 
by confronting the measured $\ttbar$ production cross section at 7 and 8 
TeV~\cite{Chatrchyan:2012bra,Chatrchyan:2013faa,CMS-PAS-TOP-13-004}. 
The obtained combined $\mtp$ results are also listed 
($\sqrt{s} =$ 7 + $\sqrt{s} =$ 8 TeV).}  
%with predictions employing different PDF sets.
%Obtained $\mtp$ at NNLO+NNLL extracted by confronting the measured
%$\sigma_{\ttbar}$ at 7 and 8 TeV~\cite{Chatrchyan:2012bra,Chatrchyan:2013faa,CMS-PAS-TOP-13-004}, with predictions employing different PDF sets.}
\label{tab:mpole}
\tabcolsep7pt\begin{tabular}{lccc}
\hline
 PDF      & $\mtp$ ($\sqrt{s} =$ 7 TeV) [GeV]& $\mtp$ ($\sqrt{s} =$ 8 TeV) [GeV]   &  $\mtp$ ($\sqrt{s} =$ 7 + $\sqrt{s} =$ 8 TeV) [GeV]\\
\hline
 NNPDF3.0~\cite{Ball:2014uwa} & 173.4 $\pm^{2.0}_{2.0}$ & 173.9 $\pm^{1.9}_{2.0}$ & 173.6 $\pm^{+1.7}_{1.8}$\\
 MMHT2014~\cite{Harland-Lang:2014zoa} & 173.7 $\pm^{2.0}_{2.1}$ & 174.2 $\pm^{1.9}_{2.2}$ & 173.9 $\pm^{+1.8}_{1.9}$\\
 CT14~\cite{Dulat:2015mca}     & 173.9 $\pm^{2.3}_{2.4}$ & 174.3 $\pm^{2.2}_{2.4}$ & 174.1$\pm^{+2.1}_{2.2}$\\
\hline
\end{tabular}
\end{table}
%
%\begin{table}[ht]
%\caption{Combined $\mtp$ at NNLO+NNLL extracted by confronting the measured
%$\sigma_{\ttbar}$ with predictions employing different PDF sets.}
%\label{tab:mpoletop}
%\tabcolsep7pt\begin{tabular}{lc}
%\hline
% PDF       & $\mtp$ ($\sqrt{s} =$ 7 + $\sqrt{s} =$ 8 TeV) [GeV]   \\
%\hline
% NNPDF3.0& 173.6 $\pm^{+1.7}_{1.8}$ \\
% MMHT2014& 173.9 $\pm^{+1.8}_{1.9}$ \\
% CT14    & 174.1$\pm^{+2.1}_{2.2}$  \\
%\hline
%\end{tabular}
%\end{table}
Figure~\ref{fig:1} shows the combined likelihood of the measured and predicted 
dependence of the $\ttbar$ production cross section on $\mtp$ for 7 
(left plot) and 8 TeV (right plot).

\begin{figure}[ht]
%  \centerline{\includegraphics[width=250pt]{CMS-PAS-TOP-14-014-Figure-008.pdf}}
  \centerline{\includegraphics[width=250pt]{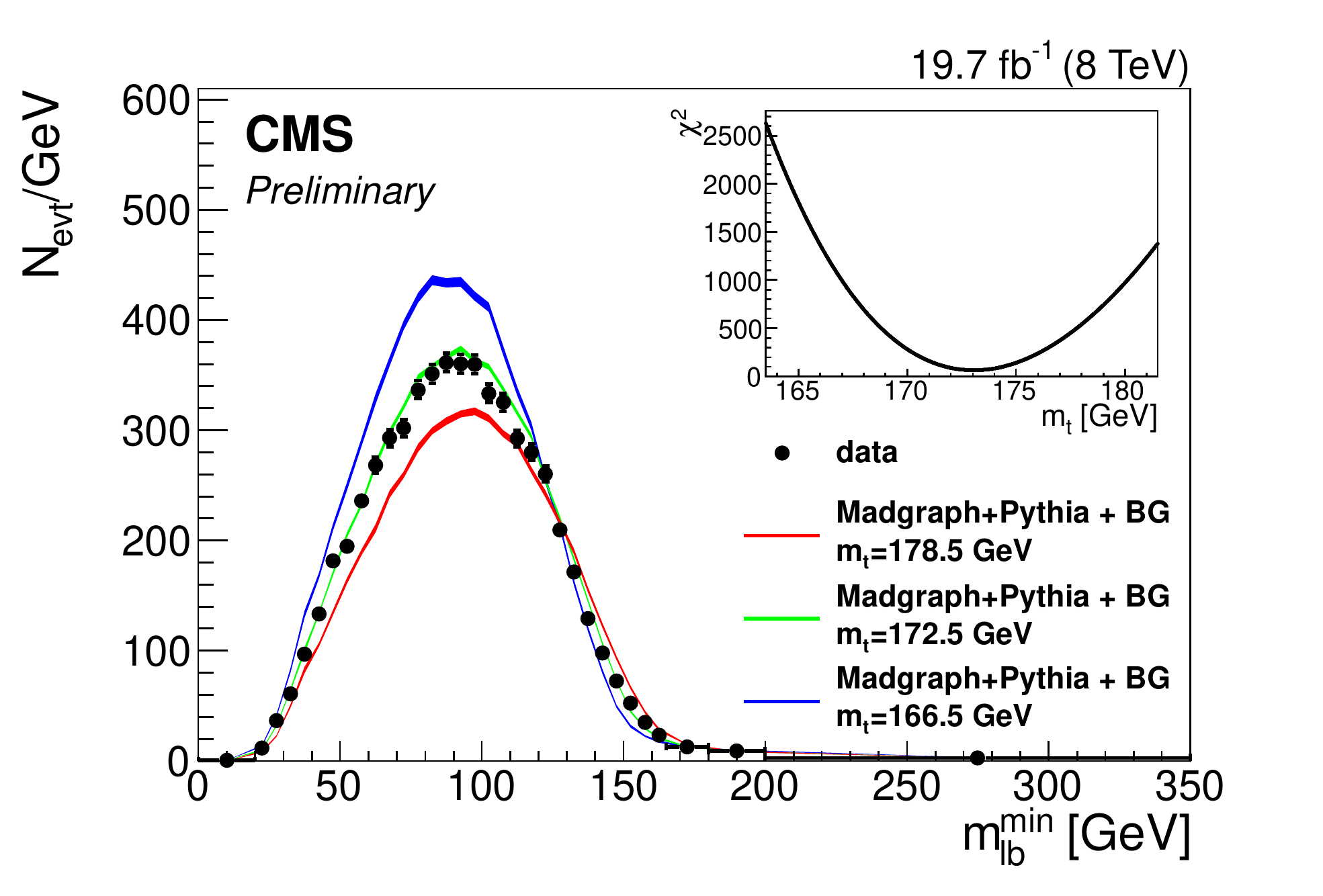}}
  \caption{Normalized event yields obtained by CMS~\cite{CMS-PAS-TOP-14-014}, 
     for $\ttbar$ production at the LHC at $\sqrt{s} =$ 8 TeV, presented as 
     a function of $\mmlb$. The bullets 
     are the experimental data points and the error bars indicate their 
     statistical uncertainties. The inset shows the $\chi^2$ distribution as a 
     function of $\mt$ as determined from the fit of the simulation to the 
     shape of the data.}
%Plot shows the shape of $\mmlb$ obtained in CMS. Right plot shows 
%   comparison of $\mtp$ values determined from ATLAS $\mtp$ measurements and 
%   previous measurements. } 
\label{fig:2}
\end{figure}

In another measurement at $\sqrt{s} =$ 8 TeV CMS~\cite{CMS-PAS-TOP-14-014}
uses a folding technique to map fixed order QCD calculations
depending on $\mtp$ as implemented in the Monte Carlo for
Femtobarn calculation MCFM~\cite{Campbell:2010ff}, 
to predict the shape in $\mmlb$. 
The top quark decay chain considered in this analysis is 
$t\to W b$ followed by $W \to l\nu$. Neglecting both leptons and 
b-quark masses, at leading order the quantity $\mlb$ is directly related 
to $\mt$ and the mass of the W boson, $m_W$, as follows: 
$\mlb^2 = \frac{\mt^2 - m_{w}^2}{2}(1 - \cos\theta_{lb})$. 
Here, $\theta_{lb}$ is the opening angle between the lepton and the b quark 
in the W-boson rest frame. The distribution of $\mlb$ has an end point at 
$max(\mlb) \sim \sqrt{\mt^2 - m_{W'}^2}$, i.e. around
153 GeV for a top-quark mass of 173 GeV.
In the analysis, $\mlb$ is reconstructed by choosing the permutation 
that minimizes the value of $\mlb$ in each event and only the b-jet candidate 
with the highest transverse momentum $p_{\perp}$ is considered together with
both leptons ($e$ and $\mu$). Only one top quark in each event is used. 
In this particular definition, the combination yielding the smallest 
$\mlb$ in the event is kept, and referred to as $\mmlb$,
shown in Figure~\ref{fig:2}. The response matrices in $\mmlb$ are 
obtained from fully simulated events obtained 
using the matrix element generator MADGRAPH 5.1.5.11~\cite{Alwall:2011uj}
with MADSPIN~\cite{Artoisenet:2012st} for the decay of heavy resonances,
PYTHIA 6.426~\cite{Sjostrand:2006za} for parton showering; 
%and using the Z2* tune~\cite{Chatrchyan:2011id}.
%using MC techniques based on the matrix element (ME) generator 
%MADGRAPH 5.1.5.11~\cite{Alwall:2011uj} interfaced with 
%MADSPIN~\cite{Artoisenet:2012st} for the decay of heavy resonances, 
%PYTHIA 6.426~\cite{Sjostrand:2006za} for parton showering 
%and hadronization using the Z2* tune~\cite{Chatrchyan:2011id}.
the MC events have been passed through a full simulation of the CMS detector 
based on GEANT~\cite{Agostinelli:2002hh} 
(combination called MADGRAPH + PYTHIA + GEANT).
%fully simulated MADGRAPH+PYTHIA+GEANT4~\cite{Alwall:2011uj,Sjostrand:2006za,Agostinelli:2002hh} events. 
%This approach leads to
By using the information on the rate of events alone a value of 
$\mt = 171.4 \pm 0.4_{stat} \pm 1.0_{syst}$ GeV is measured.
Combining the results obtained using rate+$\mmlb$ shape fits one is 
able to extract $\mt = 173.1^{1.9}_{1.8}$ GeV.
These results can be compared to the mass extraction from the same dataset via 
the total cross-section calculated at NNLO.
% previously mentioned.
%The comparison demonstrates the advantage of shape-based over total 
%cross-section based methods.
% Figure
\subsubsection{ATLAS Top-Quark Pole Mass Measurements}
\begin{figure}[ht]
%  \centerline{\includegraphics[width=250pt]{ATLASarXiv1406-5375fig-08a.pdf}}
  \centerline{\includegraphics[width=250pt]{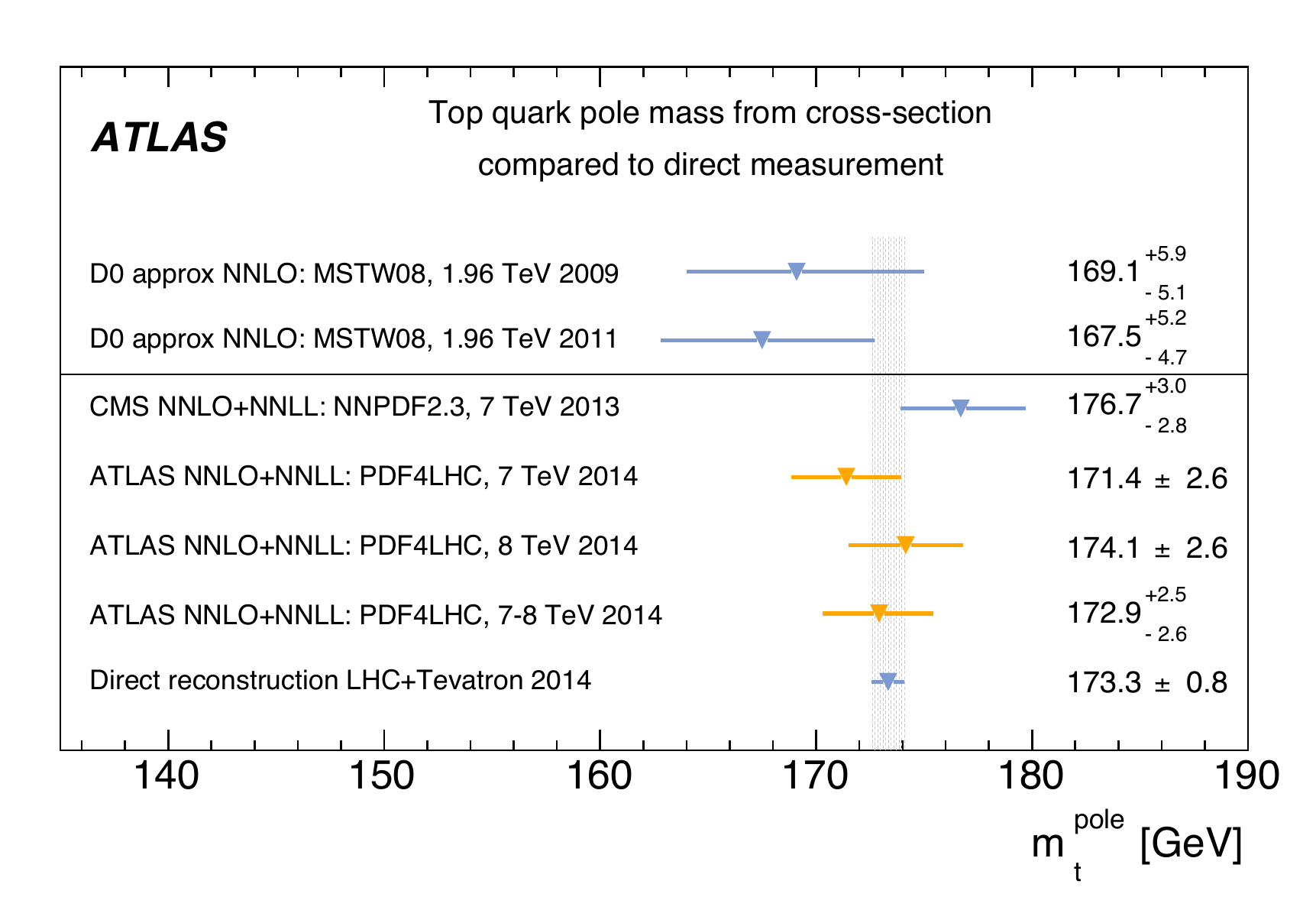}}
  \caption{Comparison of $\mtp$ values determined from 
   ATLAS and previous measurements~\cite{Aad:2014kva}. } 
\label{fig:3}
\end{figure}

ATLAS also extracts $\mtp$ at NNLO+NNLL 
by confronting the measured production cross section $\sigma_{\ttbar}$ 
at 7 and 8 TeV with predictions employing different PDF 
sets~\cite{Aad:2014kva}: CT10 NLO~\cite{Lai:2010vv}, 
MSTW 2008 68\% CL NLO~\cite{Martin:2009iq}, and NNPDF 2.3 NLO~\cite{Ball:2012cx}. 
The extraction of $\mtp$ is performed by maximizing a Bayesian likehood 
function separately for each PDF set and centre-of-mass energy to give  
$\mtp$ values shown in Table~\ref{tab:mpoletop1}. 
\begin{table}[ht]
\caption{Measurements performed by ATLAS of $\mtp$ at NNLO+NNLL 
extracted by confronting the measured production cross section 
$\sigma_{\ttbar}$ with predictions employing different PDF sets.}
\label{tab:mpoletop1}
\tabcolsep7pt\begin{tabular}{lcc}
\hline
  PDF & ATLAS $\mtp$ $\sqrt{s} =7$ TeV [GeV] & ATLAS $\mtp$ $\sqrt{s} =8$ TeV   [GeV]    \\
\hline
 C10 NNLOCT10 NLO~\cite{Lai:2010vv}      & 171.4 $\pm$ 2.6    & 174.1 $\pm$ 2.6\\
 MSTW 68 \% NNLO~\cite{Martin:2009iq} & 171.2 $\pm$ 2.4    & 174.0 $\pm$ 2.5\\
 NNPDF2.3 5f FFN~\cite{Ball:2012cx} & $171.3^{+2.2}_{-2.3}$& 174.2 $\pm$ 2.4 \\
\hline
\end{tabular}
\end{table}
Finally $\mtp$ is extracted from the combined $\sqrt{s} =$ 7
and $\sqrt{s} =$ 8 TeV dataset. The resulting value using the
envelope of all three considered PDF sets is $\mtp = 172.9^{+2.5}_{-2.6}$ GeV.
The results are shown in Figure~\ref{fig:3}, together with previous
determinations using similar techniques from D0~\cite{Abazov:2009ae}
and CMS~\cite{Chatrchyan:2013haa}. 
% Figure
All extracted values are consistent with the
average of measurements from kinematic reconstruction of
$\ttbar$ events of $173.34 \pm 0.76$ GeV~\cite{ATLAS:2014wva}, showing good 
compatibility of top-quark masses extracted using very different
techniques and assumptions.
\subsection{Measurements of Top-Quark Pole Mass in $\ttbar\,+$ 1-Jet Events}
\begin{figure}[ht]
%  \centerline{\includegraphics[width=240pt]{1507-01769fig-04.pdf}
%              \includegraphics[width=280pt]{ATLAS1507-01769figaux_09.pdf}}
  \centerline{\includegraphics[width=240pt]{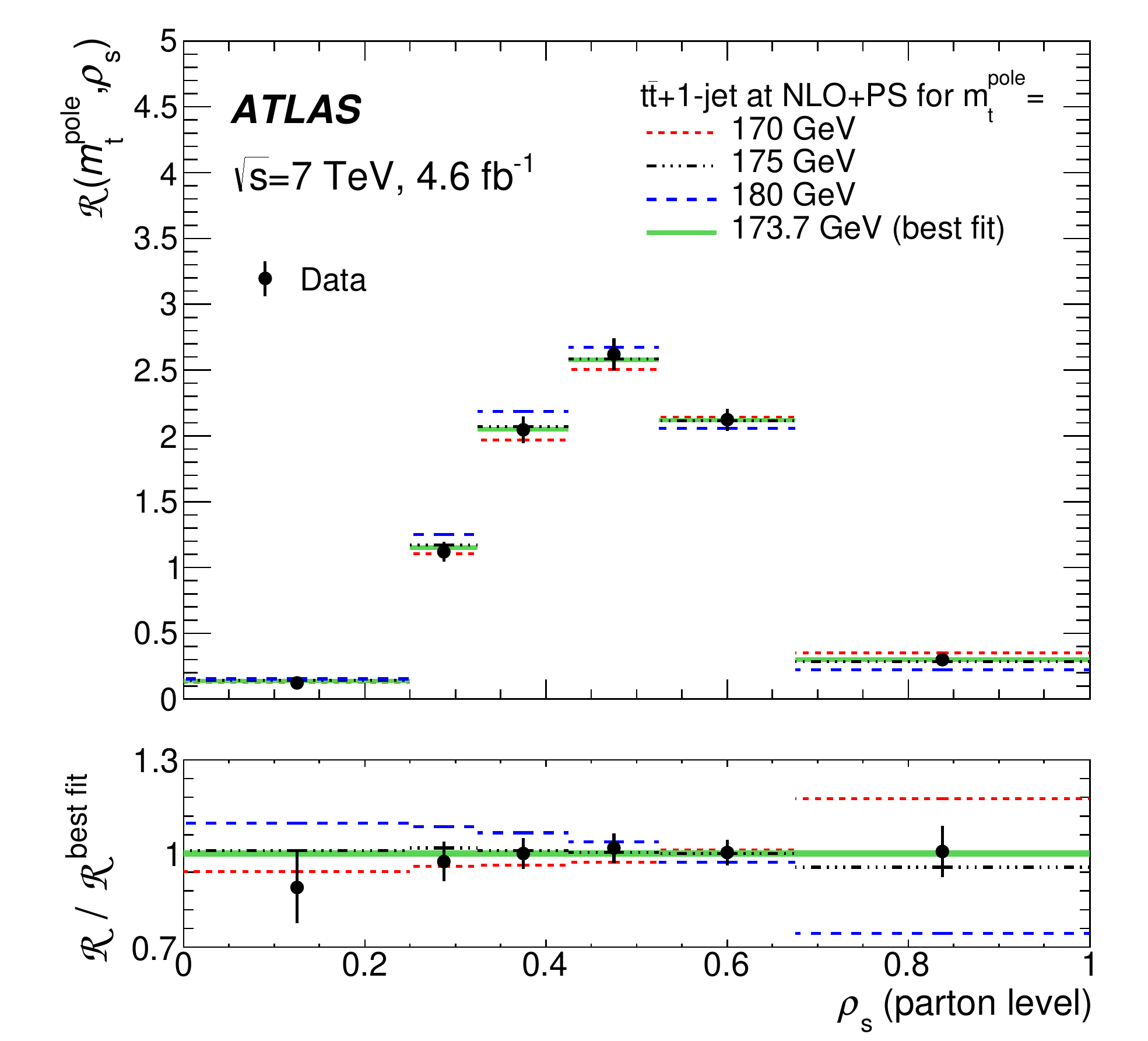}
              \includegraphics[width=280pt]{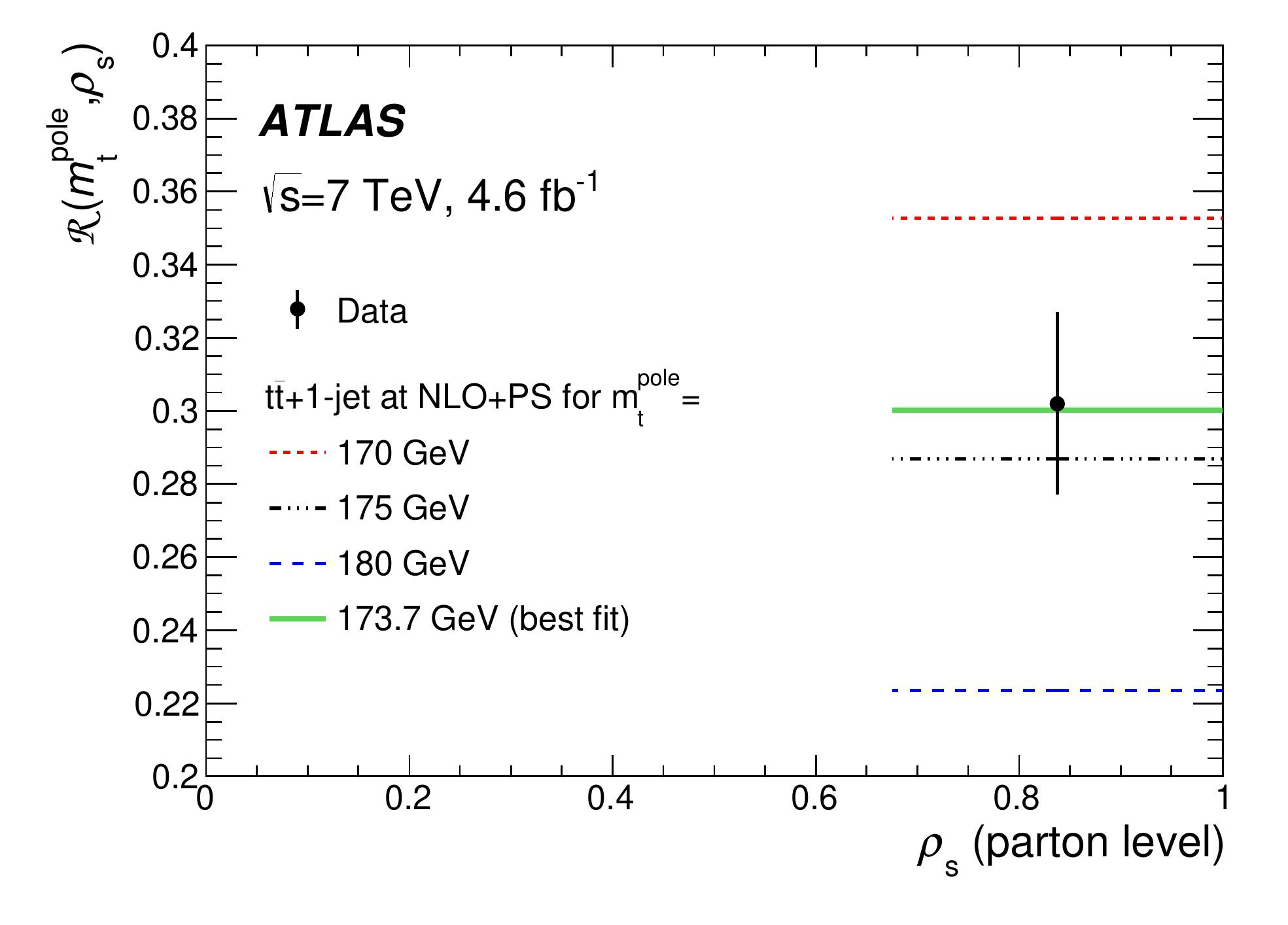}}
  \caption{Unfolded ${\cal R}(\mtp,\rho_s)$ distribution as 
   measured by ATLAS~\cite{Aad:2015waa} (left). The predictions of the 
   $\ttbar$ + 1-jet calculation at NLO+PS using three different masses 
  ($\mtp =$ 170, 175 and 180 GeV) are shown with the result of the best 
   fit to the data, $\mtp =$ 173.7 $\pm$ 1.5 (stat.) GeV.  
   The value of the most sensitive interval of the ${\cal R}$-distribution 
   $\rho_s >$ 0.65~\cite{Aad:2015waa} (right). The black point corresponds 
   to the data. 
   The shaded area indicates the statistical uncertainty of this bin.}
%This value corresponds to data corrected for detector and hadronization effects after the background subtraction.}
\label{fig:4}
\end{figure}

The normalized differential cross section for $\ttbar$ production in 
association with at least 1-jet is studied as a function of the inverse 
of the invariant mass of the $\ttbar\,+$ 1-jet system.
This distribution is used by the ATLAS experiment~\cite{Aad:2015waa} 
for a precise determination of $\mtp$.
% since gluon radiation depends on the mass of the quarks.
A new observable suggested in \cite{Alioli:2013mxa} is used in this 
measurement: 
${\cal R}(\mtp,\rho_s) = \frac{1}{\ttbarjet}\frac{d\ttbarjet}{d \rho_s}(\mtp,\rho_{s})$.
The differential is taken in $\rho_s = 2 m_0 / \sqrt{s_{\ttbar j}}$, 
that is the ratio of an arbitrary mass scale in the vicinity of $\mt$, here
set to $m_0 = 170$ GeV, over the invariant $\ttbar + 1\,$jet mass.
$\ttbar$ events are selected at $\sqrt{s} = 7$ TeV in a similar
way as done for the the lepton+jets analysis~\cite{Aad:2015nba}, and
an additional central jet with $p_{\perp} > 50$ GeV is added.
An SVD unfolding~\cite{Hocker:1995kb} with a response matrix from 
POWHEG+PYTHIA+GEANT4~\cite{Sjostrand:2006za,Agostinelli:2002hh,Alioli:2010xd}
maps the measured $\rho_s$ to parton level.
The unfolded distribution of ${\cal R}(\mtp,\rho_s)$ is shown in 
Figure~\ref{fig:4} (left).
The measurement of $\mtp =$ 173.7 $\pm 1.5_{stat} \pm 1.4_{syst}$ GeV is 
then obtained in a $\chi^2$-fit to $0.25 < \rho_s < 1$ with $\rho_s > 0.675$ 
being the most sensitive bin, as shown in Figure~\ref{fig:4} (right).
\section{CONCLUSIONS}

Measurements of $\mtp$ using alternative methods have been performed
by both the ATLAS and CMS experiments using data collected
at $\sqrt{s} =$ 7 and 8 TeV at LHC.
The latest CMS~\cite{CMS-PAS-TOP-13-004} results obtained using
di-lepton $\ttbar$ events at 7 and the 8 TeV give a
$\mtp =$ 173.6 $\pm^{+1.7}_{1.8}$ GeV.
The normalized differential cross section for $\ttbar$ production in
association with at least 1-jet studied by the
ATLAS experiment~\cite{Aad:2015waa} at 7 TeV give a
measurement of $\mtp =$ 173.7 $\pm 1.5_{stat} \pm 1.4_{syst}$ GeV.
All the extracted values of $\mtp$ are consistent with
$\mt$ measurements obtained using standard kinematic reconstruction of
$\ttbar$ events.
\section{ACKNOWLEDGMENTS}
I would like to thank the ATLAS and CMS Collaborations for giving me the 
opportunity to give this talk at this conference and the top-quark groups 
conveners of both experiments for providing me the material presented
here. 
%

% References
%
\nocite{*}
\bibliographystyle{aipnum-cp}%
\bibliography{TeresaBarillariLHCP2015}%
\end{document}